\documentclass{elsart}
\usepackage{amsmath,amssymb}

\usepackage{graphicx}
\usepackage{bm}
\newcommand{\z}{{\bf z}}
\newcommand{\y}{{\bf y}}
\newcommand{\x}{{\bf x}}
\newcommand{\p}{{\bf p}}
\newcommand{\q}{{\bf q}}
\newcommand{\uu}{{\bf u}}

\begin{document}
\journal{Physics Letters A}
 
\begin{frontmatter}
 
\title{Fictitious-time wave-packet dynamics in atomic systems}
\author{Toma\v{z} Fab\v{c}i\v{c}, J\"org Main, and G\"unter Wunner}
\address{Institut f\"ur Theoretische Physik 1,
         Universit\"at Stuttgart, 70550 Stuttgart, Germany}
\maketitle

\begin{abstract}
Gaussian wave packets (GWPs) are well suited as basis functions to describe 
the time evolution of arbitrary wave functions in systems with nonsingular 
smooth potentials.  They are less so in atomic systems on account of the 
singular behavior of the Coulomb potential.  We present a time-dependent 
variational method that makes the use of GWPs possible in the description 
of propagation of quantum states also in these systems.  We do so 
by a regularization of the Coulomb potential, and by introduction of a 
fictitious time coordinate in which the evolution of an initial state can 
be calculated exactly and analytically for a pure Coulomb potential.
Therefore in perturbed atomic systems variational approximations only arise 
from those parts of the potentials which deviate from the Coulomb potential.
The method is applied to the hydrogen atom in external magnetic
and electric fields.  It can be adapted to systems with definite symmetries, 
and thus allows for a wide range of applications.
\end{abstract}

\end{frontmatter}

\section{Introduction}
Wave packets in atomic systems can be excited experimentally, e.g., with 
microwaves \cite{Bay74,Gal88,Mae09} or short laser pulses \cite{Jon93,Jon96}.
Theoretically, the time evolution of wave packets can be calculated accurately
by numerically solving the time-dependent Schr\"odinger equation \cite{Par86}, 
or by using approximation methods, such as semiclassical \cite{Alb99} or 
time-dependent variational \cite{Hor92} methods.
The topic of wave packet dynamics in systems with Coulomb interactions covers
a large body of problems ranging from atomic physics to physics of solid 
state, where Coulomb interaction plays an important, often crucial, role.
In many-body physics, in particular, in the physics of solid state, 
theoretical methods well suited for studying the effects stemming from 
Coulomb interactions are still lacking.
The majority of the available methods, e.g., the method of pseudopotentials 
in atomic physics and the Fermi and the Luttinger liquid theories for solid 
conductors, are basically indirect and substantiated neither from the
theoretical nor from the experimental side.
For this reason they still remain, to a certain extent, disputable.
The time-dependent variational principle (TDVP) applied to Gaussian wave 
packets (GWPs) lead to exact results for the harmonic oscillator potential.
GWPs have turned out to be also well suited for describing the time evolution 
of arbitrary wave functions in smooth and nearly harmonic potentials 
\cite{Hel75,Hel76} but they are bound to fail for atomic systems because of 
the singularity of the Coulomb potential.
It is the objective of this Letter to make the GWP method applicable to the 
description of the time evolution of arbitrary quantum states also in these 
systems.

For the one-dimensional (1D) Coulomb potential, attempts already have been 
made \cite{Barnes93,Barnes94,Barnes95} to use GWPs as trial wave functions,
based on a local harmonic approximation.
For the full 3D Coulomb potential, which we will consider, a way to remove 
the singularity is, as is well known, the transformation to 4D 
Kustaanheimo-Stiefel (KS) coordinates \cite{Kus65,Stiefel71}, which converts 
the Coulomb potential into a sum of two 2D harmonic oscillator potentials, 
adapted to the use of GWPs, but also introduces an additional constraint 
on the wave functions.
The regularization implies a fictitious time coordinate.
Various approaches have been made to construct coherent states for the 
hydrogen atom \cite{Ger86,Tad98,Xu00,Una01,Polshin01} in the fictitious time 
in analogy with the coherent states of the harmonic oscillator.
These approaches construct the coherent states as the eigenstates of the 
lowering operators associated with the harmonic potential.

We will present a variant of the GWP method in coordinate
space, which describes wave packet propagation in the Coulomb problem 
{\em exactly}.
Therefore it is only deviations from the Coulomb potential which require 
a variational treatment. 
As prime examples we will apply the method to wave packet propagation
in the hydrogen atom in a magnetic field, and in crossed electric and 
magnetic fields.
Both systems have attracted considerable attention over the past decades 
because classically they exhibit a transition from regular to chaotic motion 
and thus can be used in the search for quantum signatures of chaos 
\cite{Fri89,Has89}.
If the method can be further extended to larger systems with more degrees
of freedom it will allow for a wide range of future applications in 
different branches of physics.

\section{Restricted Gaussian wave packets}
The Hamiltonian for an electron under the combined action
of the Coulomb potential and external perpendicularly 
crossed electric and magnetic fields has the form
(in atomic units, with $F_0=5.14\times 10^9$~V/cm, $B_0=2.35\times 10^5$~T)
\begin{equation}
 H_3 = \frac{1}{2}{\bf p}^2 - \frac{1}{r} + \frac{1}{2}Bl_z
     + \frac{1}{8}B^2(x^2+y^2) + Fx \; .
\label{eq:H3}
\end{equation}
Here we have assumed that the electric field is oriented along the
$x$ axis, and the magnetic field along the $z$ axis.
We regularize the singularity of the Coulomb potential by switching to KS 
coordinates ${\bf u}$ with $x=u_1u_3-u_2u_4$, $y=u_1u_4+u_2u_3$, and 
$z=\frac{1}{2}(u_1^2+u_2^2-u_3^2-u_4^2)$.
Introducing scaled coordinates and momenta
${\bf u}\to n_{\rm eff}^{1/2}{\bf u}$, ${\bf p}_u\to n_{\rm eff}^{-1/2}{\bf p}_u$
one obtains
\begin{equation}
 H\psi = \left[ \frac{1}{2}{\bf p}_u^2
  + V({\bf p}_u,{\bf u}) \right] \psi = 2n_{\rm eff}\, \psi \; ,
\label{regH4D}
\end{equation}
where the scaled potential $V$ depends on the parameters 
\begin{equation}
 \alpha \equiv -n_{\rm eff}^2E \; , \quad
 \beta \equiv n_{\rm eff}^2B \; , \quad
 \zeta \equiv n_{\rm eff}^3F \; , 
\label{eq_neff_e_feld}
\end{equation}
which can be chosen constant.
Eq.\ (\ref{regH4D}) is an eigenvalue problem for the effective quantum
number $n_{\rm eff}$, and for any quantized $n_{\rm eff}$ the energy $E$ and 
field strengths $B$ and $F$ of the physical state are obtained from 
Eq.\ (\ref{eq_neff_e_feld}).
In KS coordinates physical wave functions must fulfill the constraint
\begin{equation}
 (u_2 p_1 - u_1 p_2 - u_4 p_3 + u_3 p_4)\, \psi = 0 \; .
\label{eq_rest}
\end{equation}
For $\alpha=1/2$ and vanishing external fields ($\beta=0$, $\zeta=0$) 
Eq.\ (\ref{regH4D}) describes the 4D harmonic oscillator
with $V=\frac{1}{2}{\bf u}^2$, and $n_{\rm eff}=n$ becomes the principal 
quantum number of the field-free hydrogen atom.

Eq.\ (\ref{regH4D}) can be extended to the time-dependent 
Schr\"odinger equation in a fictitious time $\tau$ by the 
replacement $2n_{\rm eff}\to i\frac{\partial}{\partial\tau}$, viz.\
\begin{equation}
   i\frac{\partial}{\partial\tau}\psi
 = \left(\frac{1}{2}{\bf p}_u^2 + V\right)\psi = H\psi \; .
\label{eq_TVpsi}
\end{equation}
When the TDVP is used to solve Eq.\ (\ref{eq_TVpsi}) the wave function $\psi$ 
depends on a set of appropriately chosen parameters whose time-dependences
are obtained by solving ordinary differential equations.
As the regularized Hamiltonian without external fields (\ref{regH4D}) 
becomes that of a harmonic oscillator basis trial wave functions in the
form of GWPs
\begin{equation}
   g(A,\q,{\boldsymbol \pi},\gamma)
 = e^{i[(\uu-\q)A(\uu-\q)+{\boldsymbol \pi}\cdot(\uu-\q)+\gamma]}
\label{GWPhel1}
\end{equation}
with time-dependent parameters are a natural choice.
In (\ref{GWPhel1}) $A$ designates a complex symmetric $4\times 4$ 
width matrix with positive definite imaginary part, ${\boldsymbol\pi}$ 
and $\q$ are the expectation values of the momentum and position operator, 
and the phase and normalization are given by the complex scalar $\gamma$.
In KS coordinates physical wave functions must fulfill the constraint
(\ref{eq_rest}).
Inserting the ansatz (\ref{GWPhel1}) into (\ref{eq_rest}) leads to
restrictions for the admissible variational parameters, viz.\ 
$\q=0$, $\boldsymbol\pi=0$, and the special form of the width matrix
\begin{equation}
 A = \left(\begin{array}{crcr}
  a_{\mu} & 0\; & a_{x} & a_{y} \\
  0      & a_{\mu} & a_{y} & -a_{x} \\
  a_{x} & a_{y} & a_{\nu} & 0\; \\
  a_{y} & -a_{x} & 0     & a_{\nu} 
\end{array}\right) \; ,
\label{eq_A_sym}
\end{equation}
which depends only on four parameters $(a_\mu,a_\nu,a_x,a_y)$ 
\cite{Fab09a,Fab09b}.
The ``restricted Gaussian wave packets'' obeying Eq.\ (\ref{eq_rest})
are located around the origin with zero mean velocity and thus at first
glance might not appear appropriate for dynamical calculations.
However, they are the key for both the exact analytical derivation of the 
fictitious time wave packet dynamics in the field-free hydrogen atom and
the time-dependent variational approach to the perturbed atom.

The restricted GWPs are not a complete basis set for the four-dimensional
harmonic oscillator but they are in the 3D space, i.e., any physically 
allowed state can be expanded in that basis.
This can be verified by transforming the restricted GWP in KS coordinates 
back into 3D Cartesian coordinates,
\begin{subequations}
\label{eq_res_gwp}
\begin{align}
\label{eq_gwpKS}
 g({\bf y}) &= e^{i(\uu A \uu + \gamma)} \\
 &= e^{i[(a_\mu+a_\nu) r+(a_\mu-a_\nu) z  +2a_x x+2a_y y+\gamma]} \\
 &= e^{i(p_r r+ \p \cdot \x + \gamma)} \; .
\label{eq_KS_gwp_x}
\end{align}
\end{subequations}
In (\ref{eq_KS_gwp_x}) the set of parameters $(a_\mu,a_\nu,a_x,a_y,\gamma)$ 
is replaced with an equivalent set $\y=(p_r,\p,\gamma)$ with
\begin{equation}
 p_r = a_\mu+a_\nu \, , \;
 {\bf p} = (p_x,p_y,p_z) = (2a_x,2a_y,a_\mu-a_\nu) \; .
\label{eq_aparam_cart}
\end{equation}
For $p_r=0$ and real valued parameters $p_x,p_y,p_z$ the restricted GWP in 
Cartesian coordinates (\ref{eq_KS_gwp_x}) reduces to a plane wave.
Since plane waves form a complete basis we have the result that the 
restricted GWPs (\ref{eq_res_gwp}) are also complete, or even over-complete.

\section{Time-dependent variational principle}
The propagation of the wave packets is investigated by applying the TDVP.
Briefly, the TDVP of McLachlan \cite{McL64}, or equivalently the 
minimum error method \cite{Saw85}, requires to minimize the deviation between
the right-hand and the left-hand side of the time-dependent Schr\"odinger 
equation with the trial function inserted.
The quantity
\begin{equation}
 I = ||i \phi(\tau) -H \psi(\tau)||^2 \overset{!}{=} \min
\label{mem}
\end{equation}
is to be varied with respect to $\phi$ only, and then $\dot\psi\equiv\phi$ 
is chosen, i.e., for any time $\tau$ the fixed wave function $\psi(\tau)$ 
is supposed to be given and its admissible time derivative $\dot\psi(\tau)$ 
is determined by the requirement to minimize $I$.
As trial functions we consider superpositions of $N$ restricted GWPs 
(\ref{eq_gwpKS}), i.e.,
\begin{equation}
 \psi(\tau) = \psi(\z(\tau)) = \sum_{k=1}^N g({\bf y}^k)
 \equiv \sum_{k=1}^N g^k \; ,
\label{eq_psi_tau}
\end{equation}
which are parameterized by a set of $5N$ time-dependent complex parameters
$\z=(\y^k,k=1,\dots,N)$ [instead of $15N$ complex parameters when using the 
most general superposition of Gaussian wave packets (\ref{GWPhel1}) in 
4D coordinate space].
The equations of motion for the variational parameters $\z(\tau)$ are 
obtained as
\begin{subequations}
\label{eq_gwp_reg_eq_motion_var_both}
\begin{align}
\label{eq_gwp_reg_eq_motion_var}
 \dot A^k &=  -2(A^k)^2-\frac{1}{2}V_2^k \; , \\
 \dot \gamma^k &= i\, {\rm tr}\,A^k -v_0^k \; ,
\end{align}
\end{subequations}
where we have introduced the time-dependent scalars $v_0^k$ and matrices 
$V_2^k$, which induce couplings between the restricted GWPs.
Since the special structure of the matrices $A^k$ in Eq.\ (\ref{eq_A_sym}) is 
maintained in the squared matrices $(A^k)^2$, that structure carries over to
the $ 4\times 4$ complex symmetric matrices $V_2^k$.
Therefore, they have only four independent coefficients 
$(V_\mu^k,V_\nu^k,V_x^k,V_y^k)$, in the notation of Eq.\ (\ref{eq_A_sym}).
The parameters $v_0^k$ and $V_2^k$ are calculated at each time step by 
solving a $5N$ dimensional set of linear equations.
All integrals required for the setup of that linear system have the form 
$\langle g^l|f({\bf u},{\bf p}_u)|g^k\rangle$, with $f({\bf u},{\bf p}_u)$ 
a polynomial in the KS coordinates and momenta, and can be calculated
analytically \cite{Fab09b}.

For the field-free hydrogen atom one finds
$v_0^k=0$ and $V_2^k={\boldsymbol 1}$, i.e., the equations of motion
(\ref{eq_gwp_reg_eq_motion_var_both}) simplify to the uncoupled equations
$\dot A=-2A^2-\frac{1}{2}{\boldsymbol 1}$,
$\dot\gamma=i\,{\rm tr}\, A$ for the parameters of each each basis state.
These equations can be solved analytically \cite{Fab09a} and yield for
the time evolution of the restricted GWP the explicit form
\begin{equation}
 g(\tau)  = \frac{1}{\mathcal N(\tau)}
  \exp\left\{i\,\frac{\mathcal Z(\tau)}{\mathcal N(\tau)} \right\} \; ,
\label{eq_GWP_tau}
\end{equation}
with
\begin{subequations}
\begin{align}
{\mathcal Z(\tau)} &= \p^0\cdot \x +p_r^0 r\cos2\tau
  + \frac{r}{2}[(p_r^0)^2-(\p^0)^2-1]\sin2\tau \; ,\\
{\mathcal N(\tau)} &= 1 + [(p_r^0)^2-(\p^0)^2](1-\cos2\tau) + p_r^0\sin2\tau\; ,
\end{align}
\end{subequations}
and where $p_r^0$ and $\p^0$ are the parameters (\ref{eq_aparam_cart}) of 
the initial GWP at time $\tau=0$.
This is an important result for the field-free hydrogen atom:
The time evolution of a restricted GWP (\ref{eq_KS_gwp_x}) can be calculated
analytically, and takes the compact form (\ref{eq_GWP_tau}), which is a 
periodic function of the fictitious time $\tau$ with period $\pi$.
In the physical time wave packets disperse in the hydrogen atom.
By contrast, the wave packets in the fictitious time show an oscillating 
behavior, with no long-time dispersion in $\tau$.

\section{Gaussian wave packet dynamics}
We now investigate the propagation of 3D Gaussian wave packets which are 
localized around a given point $\x_0$ with width $\sigma$ in coordinate space,
and around $\p_0$ in momentum space.
As mentioned above any physical state can be expressed in terms of the 
complete basis set of the restricted GWPs (\ref{eq_res_gwp}).
The Fourier decomposition of the initial 3D GWP has the form
\begin{eqnarray}
 \psi &=& (2\pi \sigma^2)^{-3/4}
 \exp\left\{-\frac{(\x-\x_0)^2}{4\sigma^2}+i\p_0\cdot(\x-\x_0)\right\} 
 \nonumber \\
 &=& \left(\frac{\sigma^2}{2\pi^3}\right)^{3/4}
  \int d^3p\,e^{-\sigma^2(\p-\p_0)^2-i\p\cdot\x_0}\, g({\bf y}) \; ,
\label{invFTpsi2}
\end{eqnarray}
where the $g({\bf y})$ are the restricted GWPs (\ref{eq_KS_gwp_x}) for 
the set of parameters ${\bf y}$ given as $(p_r=0,\p,\gamma=0)$.

In numerical computations it is convenient to approximate the initial 3D 
Gaussian wave packet by a finite number of restricted GWPs rather than using 
the integral representation (\ref{invFTpsi2}).
This is most efficiently achieved by evaluating the integral in 
(\ref{invFTpsi2}) by a Monte Carlo method using importance sampling of 
the momenta.
The initial wave packet then reads
\begin{equation}
 \psi = (2\pi\sigma^2)^{-3/4} \frac{1}{N}
 \sum_{k=1}^N g\left({\bf y}^k\right) \, e^{-i\,\p^k\cdot\x_0} \; ,
\label{psi_MC}
\end{equation}
with ${\bf y}^k=(i\epsilon,\p^k-i\epsilon\x_0/|\x_0|,0)$, and the $\p^k$
distributed randomly according to the normalized Gaussian weight function 
$w(\p)=(\sigma^2/\pi)^{3/2}\exp\{-\sigma^2(\p-\p_0)^2\}$.
A small $\epsilon>0$ has been introduced for damping of the restricted GWPs
at large radii $r$, which is convenient in numerical computations.
The wave function $\psi$ in Eq.\ (\ref{psi_MC}) is an approximation
to the 3D Gaussian wave packet (\ref{invFTpsi2}), and the accuracy
depends on how many restricted GWPs are included.
However, it is important to note that a localized wave packet can be 
described even with a rather low number $N$ of restricted GWPs.
The time propagation of an initial state (\ref{psi_MC}) in the fictitious
time $\tau$ is now obtained exactly and fully analytically by replacing
the initial restricted GWPs $g(\y^k)$ in Eq.\ (\ref{psi_MC}) with the
corresponding time-dependent solutions (\ref{eq_GWP_tau}).
Results for the wave packet propagation in the field-free hydrogen atom
are given elsewhere \cite{Fab09a}.

Here we present example calculations for the hydrogen atom in external
fields.
In crossed electric and magnetic fields the propagation of 3D GWPs is 
computed for the time-dependent Schr\"odinger equation (\ref{regH4D}) 
with parameters $\alpha = 0.5$, $\beta = 0.05$, and $\zeta = 0.01$ in 
Eq.\ (\ref{eq_neff_e_feld}).
The choice of an appropriate initial state $\psi(0)$ is very important
for the successful application of the TDVP.
We achieved optimal results by choosing a 3D initial Gaussian wave packet
in physical Cartesian coordinates as given in (\ref{invFTpsi2}).
The external fields lead to couplings between the basis states and the
time-dependence of the variational parameters must be determined by
the numerical integration of Eq.\ (\ref{eq_gwp_reg_eq_motion_var_both}).
For better numerical performance we resort to the TDVP with constraints 
\cite{Fab08}.

\begin{figure}
\begin{center}
\includegraphics[width=0.85\columnwidth]{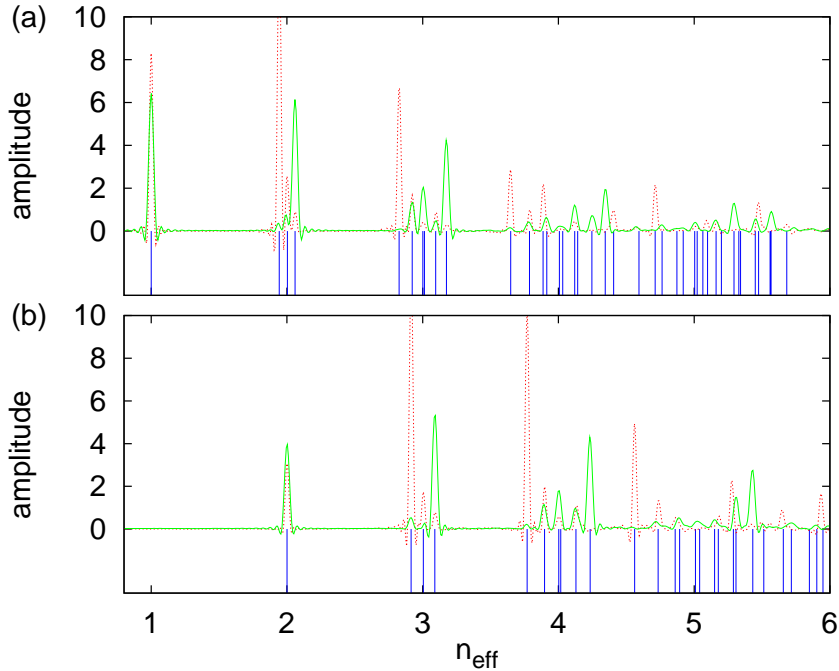}
\end{center}
\caption{
  Spectra with (a) even and (b) odd $z$ parity of the Hamiltonian
  (\ref{regH4D}) with $\alpha=0.5$, $\beta=0.05$, $\zeta = 0.01$ obtained
  from the propagation of two different 3D GWPs.  Green and red line:
  $\x_0=(6,0,0)$, $\p_{0}=(0,\pm 1/\sqrt{2},1/\sqrt{2})$, respectively.
  The eigenvalues are extracted from the autocorrelation function by Fourier
  transform.  The peak positions agree very well with the numerically exact
  eigenvalues of the effective quantum number marked by blue lines.}
\label{fig1}
\end{figure}
Once a time-dependent wave packet (\ref{eq_psi_tau}) is determined the 
eigenvalues $n_{\rm eff}$ of the stationary Schr\"odinger equation 
(\ref{regH4D}) and thus a quantum spectrum of the hydrogen atom in 
external fields can be obtained by frequency analysis of the time signal
\begin{equation}
 C(\tau) = \langle\psi(0)|\psi(\tau)\rangle 
 = \sum_j c_j e^{-2in_{\rm eff}^{(j)}\tau} \; ,
\label{eq:C_tau}
\end{equation}
with the amplitudes $c_j$ depending on the choice of the initial wave packet.
In perpendicularly crossed fields the $z$ parity is conserved.
Spectra with even and odd $z$ parity obtained from the Fourier transforms 
of the autocorrelation functions 
$C^\pm(\tau)=\langle\psi^\pm(0)|\psi^\pm(\tau)\rangle$ of the parity
projected wave packets are shown in Fig.\ \ref{fig1}.
The green and red lines result from the propagation of two different initial
3D GWPs with $\sigma=3.5$, $\epsilon = 0.15$, the same initial position 
but different initial mean momenta.
$N=41$ and $N=31$ basis states were coupled in the calculations.
The line widths, i.e., the resolution of the spectra is determined by the
length of the time signal $\tau_{\max}$.
The eigenvalues obtained by numerically exact diagonalizations of the
stationary Hamiltonian (\ref{regH4D}) are shown by the blue lines.
The line-by-line comparison shows very good agreement between the exact 
spectrum and the results obtained from the wave packet propagation.  
The amplitudes of levels indicate the excitation strengths of states with
higher or lower angular momentum $l_z$ by the two initial wave packets
rotating clockwise or anticlockwise around the $z$-axis.

\begin{figure}
\begin{center}
\includegraphics[width=0.85\columnwidth]{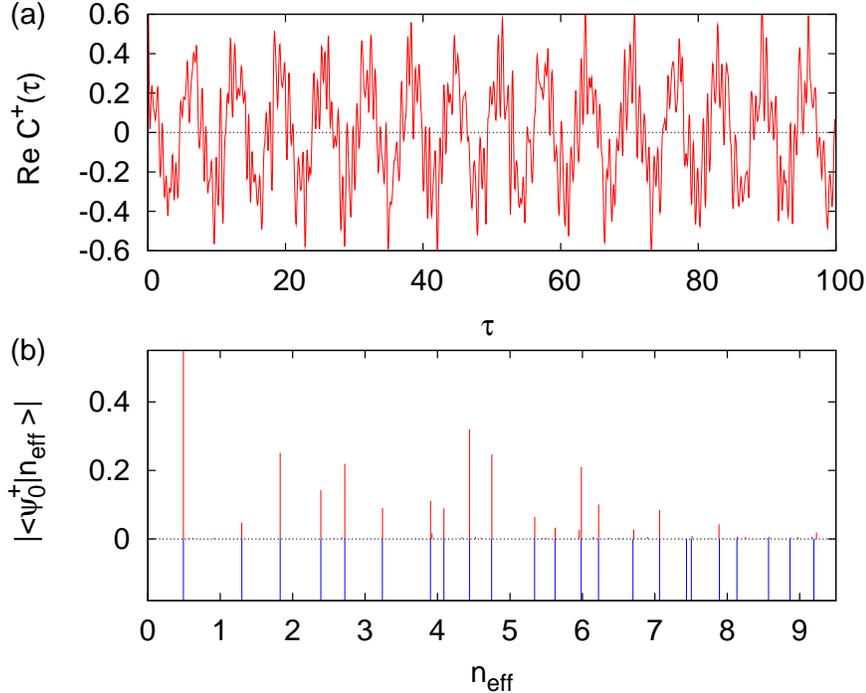}
\end{center}
\caption{
  (a) Real part of the even $z$-parity autocorrelation function and
  (b) spectrum of the diamagnetic hydrogen atom extracted from the signal
  $C^+(\tau)$ by harmonic inversion at the field-free ionization threshold
  $E=0$, $\beta=n_{\rm eff}^2B=0.5$.  For comparison the exact eigenvalues
  in (b) are marked by blue lines.  They are in excellent agreement with the
  variational results (red lines).}
\label{fig2}
\end{figure}
The method presented can be especially adapted to systems with, e.g., 
cylindrical or spherical symmetries.
For the hydrogen atom in a magnetic field we consider the very challenging
regime around the field-free ionization threshold $E=0$ where the Coulomb and
the Lorentz force are of comparable strength, resulting in a fully chaotic
classical dynamics.
The real part of the even $z$-parity autocorrelation function is shown
in Fig.\ \ref{fig2}(a).
A number of $N=90$ basis states was used in the computation.
The eigenvalues are extracted from the signal $C^+(\tau)$ by the high-resolution
harmonic inversion method \cite{Bel00} and drawn in  Fig.\ \ref{fig2}(b).
The agreement between the eigenvalues computed variationally (red lines) and 
the numerically exact results (blue lines) is very good.
Some lines are lacking in the variational computation because of a
nearly zero overlap of the respective eigenstates $|n_{\rm eff}\rangle$ 
and the initial GWP, but can be revealed by choosing different initial GWPs.

\section{Conclusion}
In this Letter we have extended the Gaussian wave packet method in such way 
that it can also be applied to quantum systems with singular Coulomb potentials.
We have shown that the evolution in fitctious time can be calculated 
analytically in the pure quantum  Coulomb problem.
Therefore in applying the time-dependent variational principle to the 
description of time evolution of wave packets in perturbed atomic systems 
approximations arise only from the non-Coulombic parts of the potentials.
The method can  be adapted to special symmetries, such as axisymmetric or 
spherical, and opens the way to a wide range of applications in systems 
with Coulomb potentials.


\begin{thebibliography}{10}
\expandafter\ifx\csname url\endcsname\relax
  \def\url#1{\texttt{#1}}\fi
\expandafter\ifx\csname urlprefix\endcsname\relax\def\urlprefix{URL }\fi

\bibitem{Bay74}
J.~E. Bayfield, P.~M. Koch, Multiphoton ionization of highly excited hydrogen
  atoms, Phys. Rev. Lett. 33 (1974) 258--261.

\bibitem{Gal88}
E.~J. Galvez, B.~E. Sauer, L.~Moorman, P.~M. Koch, D.~Richards, Microwave
  ionization of {H} atoms: Breakdown of classical dynamics for high
  frequencies, Phys. Rev. Lett. 61 (1988) 2011--2014.

\bibitem{Mae09}
H.~Maeda, J.~H. Gurian, T.~F. Gallagher, Nondispersing {B}ohr wave packets,
  Phys. Rev. Lett. 102 (2009) 103001.

\bibitem{Jon93}
R.~R. Jones, D.~You, P.~H. Bucksbaum, Ionization of {R}ydberg atoms by
  subpicosecond half-cycle electromagnetic pulses, Phys. Rev. Lett. 70 (1993)
  1236--1239.

\bibitem{Jon96}
R.~R. Jones, Creating and probing electronic wave packets using half-cycle
  pulses, Phys. Rev. Lett. 76 (1996) 3927--3930.

\bibitem{Par86}
J.~Parker, C.~R. Stroud, Coherence and decay of {R}ydberg wave packets, Phys.
  Rev. Lett. 56 (1986) 716--719.

\bibitem{Alb99}
G.~Alber, O.~Zobay, Semiclassical interferences and catastrophes in the
  ionization of {R}ydberg atoms by half-cycle pulses, Phys. Rev. A 59 (1999)
  R3174--R3177.

\bibitem{Hor92}
M.~Horbatsch, J.~K. Liakos, Generation of harmonic radiation by hydrogen atoms
  in intense laser fields, Phys. Rev. A 45 (1992) 2019--2024.

\bibitem{Hel75}
E.~J. Heller, Time-dependent approach to semiclassical dynamics, J.\ Chem.\
  Phys. 62 (1975) 1544--1555.

\bibitem{Hel76}
E.~J. Heller, Time dependent variational approach to semiclassical dynamics,
  J.\ Chem.\ Phys. 64 (1976) 63--73.

\bibitem{Barnes93}
I.~M.~S. Barnes, M.~Nauenberg, M.~Nockleby, S.~Tomsovic, Semiclassical theory
  of quantum propagation: The {C}oulomb potential, Phys.\ Rev.\ Lett. 71 (1993)
  1961--1964.

\bibitem{Barnes94}
I.~M.~S. Barnes, M.~Nauenberg, M.~Nockleby, S.~Tomsovic, Classical orbits and
  semiclassical wavepacket propagation in the {C}oulomb potential, J. Phys. A
  27 (1994) 3299--3321.

\bibitem{Barnes95}
I.~M.~S. Barnes, Semiclassical wavepacket propagation in a hydrogen atom,
  Chaos, Solitons \& Fractals 6 (1995) 531--537.

\bibitem{Kus65}
P.~Kustaanheimo, E.~Stiefel, Perturbation theory of {K}epler motion based on
  spinor regularization, Journal f\"ur die reine und angewandte Mathematik 218
  (1965) 204--219.

\bibitem{Stiefel71}
E.~Stiefel, G.~Scheifele, Linear and Regular Celestial Mechanics, Springer,
  1971.

\bibitem{Ger86}
C.~C. Gerry, Coherent states and the {K}epler-{C}oulomb problem, Phys. Rev. A
  33 (1986) 6--11.

\bibitem{Tad98}
T.~Toyoda, S.~Wakayama, Coherent states for the {K}epler motion, Phys. Rev. A
  59 (1999) 1021--1024.

\bibitem{Xu00}
B.-W. Xu, G.-H. Ding, Nonspreading coherent states for the hydrogen atom, Phys.
  Rev. A 62 (2000) 022106.

\bibitem{Una01}
N.~Unal, Parametric time-coherent states for the hydrogen atom, Phys. Rev. A 63
  (2001) 052105.

\bibitem{Polshin01}
S.~A. Pol'shin, Coherent states for the hydrogen atom: discrete and continuous
  spectra, J.\ Phys.\ A 34 (2001) 11083--11094.

\bibitem{Fri89}
H.~Friedrich, D.~Wintgen, The hydrogen atom in a uniform magnetic field -- {A}n
  example of chaos, Phys. Rep. 183 (1989) 37--79.

\bibitem{Has89}
H.~Hasegawa, M.~Robnik, G.~Wunner, Classical and quantal chaos in the
  diamagnetic {K}epler problem, Prog. Theor. Phys. Suppl. 98 (1989) 198--286.

\bibitem{Fab09a}
T.~Fab\v{c}i\v{c}, J.~Main, G.~Wunner, Fictitious-time wave-packet dynamics: I.
  {N}ondispersive wave packets in the quantum {C}oulomb problem, Phys. Rev. A
  79 (2009) 043416.

\bibitem{Fab09b}
T.~Fab\v{c}i\v{c}, J.~Main, G.~Wunner, Fictitious-time wave-packet dynamics:
  {II.} {H}ydrogen atom in external fields, Phys. Rev. A 79 (2009) 043417.

\bibitem{McL64}
A.~D. McLachlan, A variational solution of the time-dependent {S}chrodinger
  equation, Mol. Phys. 8 (1964) 39--44.

\bibitem{Saw85}
S.-I. Sawada, R.~Heather, B.~Jackson, H.~Metiu, A strategy for time dependent
  quantum mechanical calculations using a {G}aussian wave packet representation
  of the wave function, J.\ Chem.\ Phys. 83 (1985) 3009--3027.

\bibitem{Fab08}
T.~Fab\v{c}i\v{c}, J.~Main, G.~Wunner, Time propagation of constrained coupled
  {G}aussian wave packets, J.\ Chem.\ Phys. 128 (2008) 044116.

\bibitem{Bel00}
\protect{D\v{z}}. Belki{\'c}, P.~A. Dando, J.~Main, H.~S. Taylor, Three novel
  high-resolution nonlinear methods for fast signal processing, J. Chem. Phys.
  113 (2000) 6542--6556.

\end{thebibliography}

\end{document}